\documentclass[12pt]{article}
\usepackage{amsfonts,amssymb}
\usepackage{color}
\usepackage{epic}
\usepackage{graphics}

\def\lddots{\mathinner{\mkern1mu\raise1pt\hbox{.}\mkern2mu
\raise4pt\hbox{.}\mkern2mu\raise7pt\vbox{\kern7pt\hbox{.}}\mkern1mu}}
\makeatletter
\def\numberbysection{\@addtoreset{equation}{section}
\def\theequation{\thesection.\arabic{equation}}}
\makeatother

\numberbysection

\newcommand{\be}{\begin{eqnarray}}
\newcommand{\ee}{\end{eqnarray}}
\newcommand{\non}{\nonumber}

\begin{document}

\begin{titlepage}
\vskip 0.4cm \strut\hfill \vskip 0.8cm
\begin{center}

{\bf {\Large The generalized non-linear Schr\"odinger model\\ on
the interval}}

\vspace{10mm}

{\large Anastasia Doikou,\footnote{doikou@bo.infn.it}$^{a}$ Davide
Fioravanti\footnote{fioravanti@bo.infn.it}$^{a}$ and Francesco
Ravanini\footnote{ravanini@bo.infn.it}$^{a}$}

\vspace{10mm}

{\small $^a$ University of Bologna, Physics Department, INFN Section \\
Via Irnerio 46, Bologna 40126, Italy}

\end{center}

\vfill

\begin{abstract}

The generalized $(1+1)$-D non-linear Schr\"odinger (NLS) theory
with particular integrable boundary conditions is considered. More
precisely, two distinct types of boundary conditions, known as
soliton preserving (SP) and soliton non-preserving (SNP), are
implemented into the classical $gl_N$ NLS model. Based on this
choice of boundaries the relevant conserved quantities are
computed and the corresponding equations of motion are derived. A
suitable quantum lattice version of the boundary generalized NLS
model is also investigated. The first non-trivial local integral
of motion is explicitly computed, and the spectrum and Bethe
Ansatz equations are derived for the soliton non-preserving
boundary conditions.

\end{abstract}

\vfill \baselineskip=16pt

\end{titlepage}

\section{Introduction}

After various investigations (cf. below for detailed references)
it is now well established that any integrable system (with finite
or infinite degrees of freedom) based on the higher rank algebras
$gl_N$ or ${\cal U}_q(gl_N)$ may be endowed with two distinct
types of integrable \footnote{They are defined as those boundary
conditions that preserve the integrability of the system.}
boundary conditions. These boundary conditions are known as
soliton preserving (SP), traditionally recognized in the framework
of integrable quantum spin chains (finite number of degrees of
freedom) \cite{cherednik}--\cite{masa}, and soliton non-preserving
(SNP) originally introduced in the context of classical integrable
field theories (infinite number of degrees of freedom)
\cite{durham}, and further investigated in \cite{gand, dema,
dema2}. SNP boundary conditions have been also introduced and
studied for integrable quantum lattice systems \cite{ann1},
\cite{doikousnp}--\cite{crdo}, and their quantum integrability was
first shown in \cite{doikousnp}. From the physical point of view
the SP boundary conditions oblige a particle-like excitation to
reflect to itself: no multiplet changing occurs. The SNP boundary
conditions, on the other hand, force an excitation to reflect to
its `conjugate', namely to an excitation carrying the conjugate
representation. From the algebraic perspective the two types of
boundary conditions are associated with two distinct algebras,
i.e. the reflection algebra \cite{cherednik, sklyanin} and the
twisted Yangian respectively \cite{molev, moras}.

The study of the underlying algebraic structures defined by the
Yang-Baxter and reflection equations is in general of great
consequence, both at classical and quantum level, not only for
integrable systems per se, but also for other relevant problems in
theoretical physics. For instance, in the context of the AdS/CFT
correspondence \cite{MWGKP} it is known that from the string
theory side the relevant classical integrable model is a sigma
model (see e.g. \cite{BPR, KMMZ}), that is a field theory with
infinite degrees of freedom. From the quantum gauge theory side on
the other hand the loop contributions are apparently given by
integrable quantum $1$-D lattice models with a finite number of
degrees of freedom \cite{MZ, BDS}. As a consequence, a crucial
point would be the formulation of a discrete-quantum counterpart
of the aforementioned classical sigma model. In this respect, the
knowledge of the discrete-quantum Lax operator would facilitate
the derivation of the relevant Hamiltonian, and of the other
charges in involution, as well as of the exact nested Bethe Ansatz
equations. In fact, up to date only the asymptotic forms of the
{\it would-be-exact} Bethe Ansatz equations are known (see e.g.
\cite{BDS}). Thus a rigorous derivation of these equations would
undoubtedly be of great physical significance, as proven when
corrections to the asymptotic regime are available \cite{RSS,
FFGR2}.

In the frame of classical continuum theories the SNP boundary
conditions have been primarily investigated up to now
\cite{durham}--\cite{dema2}. Therefore it is of great importance
to further analyze the other set of boundary conditions, i.e. the
SP ones within this context. In the present study we examine both
SP and SNP boundary conditions for the classical generalized NLS
model, and by using Hamiltonian methods \cite{ft} we derive the
relevant integrals of motion, and also specify the corresponding
classical equations of motion. Note that in \cite{miraso} the
quantum $gl_N$ NLS model on the half line was studied, based on
the reflection algebra (i.e. SP boundaries), primarily from the
point of view of the underlying symmetry algebras. It should be
stressed that although in most classical continuum theories the
SNP boundary conditions have been analyzed (see e.g. \cite{durham,
dema, dema2}) this is the first time that such boundary conditions
are implemented within the generalized NLS model. Here we consider
the classical NLS model as a simple example, however our main
motivation is to search for all possible boundary conditions in
other classical theories such as affine Toda field theories,
principal chiral models and others. From the viewpoint of quantum
lattice models on the other hand the extensively analyzed boundary
conditions are the SP ones, thus we focus here mostly on the SNP
case for a lattice version of NLS. In particular, we consider a
suitable lattice version of the NLS model \cite{kundu, kundubas}
with SNP boundary conditions, and we derive the exact spectrum and
the corresponding Bethe Ansatz equations.

The outline of this article is as follows: in the next section we
present the generic algebraic setting for classical models on the
full line and on the interval. More precisely we introduce the
classical Yang-Baxter equation and the underlying algebra for the
system on the full line. In the case of a system on the interval
we distinguish two types of boundary conditions based on the
classical versions of reflection algebra (SP) and twisted Yangian
(SNP). Next the NLS model on the full line is reviewed and an
explicit derivation of the local integrals of motion by solving
the auxiliary linear problem \cite{ft} is presented. In section 3
being guided by the same logic and adopting Sklyanin's formulation
\cite{sklyanin} we derive the integrals of motion of the $gl_N$
NLS system on the interval with SP and SNP boundary conditions.
Moreover, the corresponding classical equations of motion are
obtained for various boundary conditions. In addition the usual
NLS model with a reflecting impurity is investigated in the same
spirit. In section 4 a suitable lattice version of the NLS model
is investigated. After a brief review on the model with periodic
boundary conditions we deal with the model with open boundaries.
First the spectrum and Bethe Ansatz equations are derived for the
usual lattice NLS. Finally, the SNP boundary conditions are
considered for the generalized NLS system. The first non-trivial
local integral of motion is explicitly specified for particular
choice of boundary conditions, and the spectrum and Bethe Ansatz
equations are deduced for the simplest boundaries.

\section{The general setting}

The line of attack which will be adopted for the study of the
$gl_N$ NLS model with integrable boundaries is based on the
solution of the so called auxiliary linear problem \cite{ft}. It
is therefore necessary to recall at least the basics regarding
this formulation. Let $\Psi$ be a solution of the following set of
equations \be &&{\partial \Psi \over
\partial x} = {\mathbb U}(x,t, \lambda) \Psi \label{dif1}\\ &&
{\partial  \Psi \over \partial t } = {\mathbb V}(x,t,\lambda) \Psi
\label{dif2} \ee with ${\mathbb U},\ {\mathbb V}$ being in general
$n \times n$ matrices with entries functions of complex valued
fields, their derivatives, and the spectral parameter $\lambda$. The
monodromy matrix from (\ref{dif1}) may be then written as: \be
T(x,y,\lambda) = {\cal P} exp \Big \{ \int_{y}^x {\mathbb
U}(x',t,\lambda)dx' \Big \}. \label{trans} \ee The fact that $T$ is
a solution of equation (\ref{dif1}) will be extensively used
subsequently for obtaining the relevant integrals of motion.
Compatibility conditions of the two differential equation
(\ref{dif1}), (\ref{dif2}) lead to the zero curvature condition \be
\dot{{\mathbb U}} - {\mathbb V}' + \Big [{\mathbb U},\ {\mathbb V}
\Big ]=0, \label{zecu} \ee giving rise to the corresponding
classical equations of motion of the system under consideration.

There exists an alternative description, known as the  $r$ matrix
approach (Hamiltonian formulation). In this picture the underlying
classical algebra is manifest in analogy to the quantum case as
will become quite transparent later. Let us first recall this
method for a general classical integrable system on the full line.
The existence of the classical r-matrix, satisfying the classical
Yang-Baxter equation \be \Big [r_{12}(\lambda_1-\lambda_2),\
r_{13}(\lambda_1)+r_{23}(\lambda_2) \Big]+ \Big
[r_{13}(\lambda_1),\ r_{23}(\lambda_2) \Big] =0, \ee guarantees
the integrability of the classical system. Indeed, consider the
operator $T(x,y,\lambda)$ satisfying \be \Big
\{T_{1}(x,y,t,\lambda_1),\ T_{2}(x,y,t,\lambda_2) \Big \} =
\Big[r_{12}(\lambda_1-\lambda_2),\
T_1(x,y,t,\lambda_1)T_2(x,y,t,\lambda_2) \Big ]. \label{basic} \ee
Making use of the latter equation one may readily show for a
system in full line: \be \Big \{\ln tr\{T(x,y,\lambda_1)\},\ \ln
tr\{T(x,y, \lambda_2)\} \Big\}=0 \ee i.e. the system is
integrable, and the charges in involution --local integrals of
motion-- may be attained by the expansion of the object $\ln
tr\{T(x,y,\lambda)\}$, based essentially on the fact that $T$ also
satisfies (\ref{dif1}).

Our main aim here is to consider the $gl_N$ NLS model on the
interval. For this purpose we follow the line of action described
in \cite{ft}, but using Sklyanin's formulation for the system on
the interval or on the half line (see also \cite{mac} for the sine
Gordon on the half line). We briefly describe this process below
for any classical integrable system on the interval. In this case
one has to derive a modified transition matrix ${\cal T}$, based
on Sklyanin's formulation and satisfying the following Poisson
bracket algebras ${\mathbb R}$, $~{\mathbb T}$, i.e. classical
versions of the reflection algebra and twisted Yangian
respectively. It will be convenient for our purposes here to
introduce some useful notation. Let \be && r_{12}^*(\lambda)
=r_{12}(\lambda) ~~~\mbox{ for SP}, ~~~~~~r_{12}^*(\lambda) = \bar
r_{12}(\lambda) =V_{1}\ r_{12}^{t_{1}}(-\lambda)\ V_{1}
~~~\mbox{for SNP} \non\\ && \hat T(\lambda) = T^{-1}(-\lambda)
~~~\mbox{for SP}, ~~~~~~\hat T(\lambda) = V\ T^{t}(-\lambda)\ V
~~~\mbox{for SNP}  \non\\ && V =\mbox{antid}(1,\ 1, \ldots, 1 )
~~~\mbox{or} ~~~V =\mbox{antid}(i,\ -i, \dots, -i) ~~~\mbox{for
$n$ even only}\label{rep2} \ee In general $V$ can be any matrix
such that $V^2 ={\mathbb I}$, for instance $V={\mathbb I}$ is also
a possible choice (see e.g. \cite{durham}). Then the defining
relations describing the classical reflection algebra and the
twisted Yangian respectively, may be written in the following
compact form\footnote{Note that the classical versions of the
reflection equation and the twisted Yangian are provided in
general by more involved expressions for generic $r$ matrices. In
the present study we focus on $r$ matrices satisfying
$r_{12}(\lambda) =r_{21}(\lambda)$ ($\bar r_{12}(\lambda)= \bar
r_{21}(\lambda)$), and in this case (\ref{refc}),  are valid.}:
\be && \Big \{{\cal T}_1(\lambda_1),\ {\cal T}_2(\lambda_2) \Big
\} = \Big [ r_{12}(\lambda_1-\lambda_2),\ {\cal
T}_{1}(\lambda_1){\cal T}_2(\lambda_2)\Big ] \non\\ && + {\cal
T}_{1}(\lambda_1) r^*_{12}(\lambda_1+\lambda_2){\cal
T}_2(\lambda_2)- {\cal T}_{2}(\lambda_2)
r^*_{12}(\lambda_1+\lambda_2){\cal T}_1(\lambda_1). \label{refc}
\ee To construct the generating function of the integrals of
motion one also needs $c$-number representations of the algebra
${\mathbb R}$ (${\mathbb T}$) satisfying (\ref{refc}) for SP and
SNP respectively, and also \be \Big \{ K_1^{\pm}(\lambda_1),\
K_2^{\pm}(\lambda_2) \Big \}=0. \label{kso} \ee The modified
transition matrices, compatible with the corresponding algebras
${\mathbb R},\ {\mathbb T}$ are given by the following expressions
\cite{sklyanin}: \be  && {\cal T}(x,y,t,\lambda) =
T(x,y,t,\lambda)\ K^{-}(\lambda)\ \hat T(x,y,t,\lambda)
\label{reps} \ee and the generating function of the involutive
quantities is defined as \be t(x,y,t,\lambda)= tr\{
K^{+}(\lambda)\ {\cal T}(x,y,t,\lambda)\}\ee Due to (\ref{refc})
it can be shown that (\ref{ref2}) \be  \Big \{t(x,y,t,\lambda_1),\
t(x,y,t,\lambda_2) \Big \} =0, ~~~ \lambda_1,\ \lambda_2 \in
{\mathbb C}. \label{bint} \ee Technical details on the proof of
classical integrability are provided in Appendix A.

By expanding $\ln\ t(\lambda)$ in powers of $\lambda^{-1}$ one
recovers the local integrals of motion of the considered system,
and this is achieved in the subsequent sections. Among the local
integrals of motion there exist naturally the Hamiltonian, which
also provides information regarding the corresponding equations of
motion. This is in fact the formulation we are going to assume for
the NLS system on the half line, although an alternative strategy
would be to derive the modified Lax pair, compatible with the
boundary conditions chosen, and hence the associated equations of
motion (see e.g. \cite{durham}). Nonetheless, the rigorous
derivation of the modified Lax pair is essentially based on the
existence of local integrals of motion \cite{ft}, therefore the
viewpoint adopted here is arguably the most natural.

\section{Classical local integrals of motion}

The main objective in this section is to solve the auxiliary
linear problem for the generalized NLS model on the interval.
Before however we proceed to the study of the model on the
interval we shall briefly review the system on the full line. In
any case, these results will be relevant for the boundary case as
well.

\subsection{The generalized NLS on the full line}

We shall hereafter focus on the $gl_N$ NLS model. Consider the
classical $r$ matrix \be r(\lambda) = {{\mathbb P} \over \lambda}
~~~~\mbox{where} ~~~~{\mathbb P}=\sum_{i,j=1}^{N} E_{ij} \otimes
E_{ji} \label{rr} \ee ${\mathbb P}$ is the permutation operator,
and $~(E_{ij})_{kl} = \delta_{ik} \delta_{jl}$. The Lax pair for
the generalized NLS model is given by the following expressions
\cite{ft}: \be {\mathbb U} = {\mathbb U}_0 + \lambda {\mathbb
U}_1, ~~~{\mathbb V} = {\mathbb V}_0+\lambda{\mathbb V}_1
+\lambda^2 {\mathbb V}_2 \ee where (see also \cite{foku}) \be &&
{\mathbb U}_1 = {1\over 2i} (\sum_{i=1}^{N-1}E_{ii} -E_{NN}),
~~~~{\mathbb U}_0 =
\sqrt{\kappa} \sum_{i=1}^{N-1}(\bar \psi_i E_{iN} +\psi_i E_{Ni}) \non\\
&&{\mathbb V}_0 = i\kappa \sum_{i,\ j=1}^{N-1}(\bar \psi_i \psi_j
E_{ij} -|\psi_i|^2E_{NN}) -i\sqrt{\kappa}\sum_{i=1}^{N-1} (\bar
\psi_i' E_{iN} - \psi_i' E_{Ni}), \non\\ && {\mathbb V}_1=
-{\mathbb U}_{0}, ~~~{\mathbb V}_2= -{\mathbb U}_{1} \label{lax}
\ee and $\psi_i,\ \bar \psi_j$ satisfy\footnote{The Poisson
structure for the generalized NLS model is defined as: \be \Big \{
A,\ B  \Big \}=  i \sum_{i} \int_{-L}^{L} dx \Big ({\delta A \over
\delta \psi_i(x)}\ {\delta B \over \delta \bar \psi_i(x)} -
{\delta A \over \delta \bar \psi_i(x)}\ {\delta B \over \delta
\psi_i(x)}\Big )  \ee}: \be \Big \{ \psi_{i}(x),\ \psi_j(y) \Big
\} = \Big \{\bar \psi_{i}(x),\ \bar \psi_j(y) \Big \} =0, ~~~~\Big
\{\psi_{i}(x),\ \bar \psi_j(y) \Big \}= \delta_{ij}\ \delta(x-y).
\ee From the zero curvature condition (\ref{zecu}) the classical
equations of motion for the generalized NLS model are entailed
i.e. \be i{\partial \psi_{i}(x,t) \over
\partial t} = - {\partial^{2} \psi_{i}(x,t) \over
\partial^2 x}+2\kappa \sum_{j}|\psi_{j}(x,t)|^2 \psi_{i}(x,t),
~~~~i,\ j \in \{1, \ldots, N-1 \}. \label{nls} \ee It is clear
that for $N=2$ the equations of motion of the usual NLS model are
recovered.

As already mentioned to obtain the local integrals of motion of
NLS one has to expand $T$ (\ref{trans}) in powers of
$\lambda^{-1}$ \cite{ft}. Let us consider the following ansatz for
$T$ as $|\lambda| \to \infty$ \be T(x,y,\lambda) = ({\mathbb I}
+W(x, \lambda))\ \exp[Z(x,y,\lambda)]\ ({\mathbb I}
+W(y,\lambda))^{-1} \label{exp0} \ee where $W$ is off diagonal
matrix i.e. $~W = \sum_{i\neq j} W_{ij} E_{ij}$, and $Z$ is purely
diagonal $~Z = \sum_{i=1}^N Z_{ii}E_{ii}$. Also \be
Z_{ii}(\lambda) = \sum_{n=-1}^{\infty} {Z^{(n)}_{ii} \over
\lambda^{n}}, ~~~~W_{ij} = \sum_{n=1}^{\infty}{W_{ij}^{(n)} \over
\lambda^n}. \label{expa} \ee Inserting the latter expressions
(\ref{expa}) in (\ref{dif1}) one may identify the coefficients
$W_{ij}^{(n)}$ and $Z_{ii}^{(n)}$ (see Appendix B for a detailed
analysis). Notice that as $i\lambda \to \infty$ the only non
negligible contribution from $Z^{(n)}$ comes from the
$Z^{(n)}_{NN}$ term, and is given by: \be Z_{NN}^{(n)}(L, -L)= i L
\delta_{n, -1} +\sqrt{\kappa}\sum_{i=1}^{N-1}\int_{-L}^L dx\
\psi_i(x)\ W_{iN}^{(n)}(x). \label{ref1} \ee It is thus sufficient
to determine the coefficients $W_{iN}^{(n)}$ in order to extract
the relevant local integrals of motion (see also \cite{foku}).
Indeed solving (\ref{dif1}) one may easily obtain: \be &&
W_{iN}^{(1)}(x) = -i \sqrt{\kappa} \bar \psi_{i}(x),
~~~~W_{iN}^{(2)}(x)=\sqrt{\kappa} \bar \psi_i'(x) \non\\ &&
W_{iN}^{(3)}(x) = i\sqrt{\kappa}\bar \psi_i''(x) -i
\kappa^{{3\over 2}} \sum_{k} |\psi_{k}(x)|^2\bar \psi_i(x),
~~\ldots. \label{ref2} \ee From the latter formulae (\ref{ref2})
and taking into account (\ref{exp0}), (\ref{ref1}) the local
integrals of motion of NLS may be readily extracted from $\ln\ tr
T(L, -L, \lambda)$, i.e. \be && I_1 = -i\kappa \int_{-L}^{L} dx\
\sum_{i=1}^{N-1}\psi_i(x) \bar \psi_i(x) , \non\\ && I_2 =-{\kappa
\over 2} \int_{-L}^{L} dx\ \sum_{i=1}^{N-1} \Big (\bar \psi_i(x)
\psi'_i(x)- \psi_i(x) \bar \psi'_i(x) \Big ), \non\\ && I_3=
-i\kappa \int_{-L}^{L} dx\ \sum_{i=1}^{N-1} \Big (\kappa
|\psi_i(x)|^2 \sum_k|\psi_k(x)|^2  +\psi_i'(x) \bar \psi_i'(x)
\Big ), ~~\ldots  \label{first} \ee The corresponding familiar
quantities for the generalized NLS are given by: \be {\cal
N}=-{I_1 \over i \kappa},~~~~{\cal P}=-{I_2 \over i \kappa},
~~~~{\cal H}=-{I_3 \over i \kappa}, \label{clas}\ee and apparently
\be \{{\cal H},\ {\cal P}\}=\{{\cal H},\ {\cal N}\}=\{ {\cal N},\
{\cal P} \}=0. \ee Again in the special case where $N=2$ the well
known local integrals of motion for the usual NLS model on the
full line are recovered.

\subsection{The generalized NLS on the interval}

After the review on the NLS on the full line we can come to our
main concern, which is the evaluation of the integrals of motion
after implementing integrable boundary conditions. We shall
investigate subsequently both SP and SNP boundary conditions.

\subsubsection{SP boundary conditions}

Let us  first consider the NLS model with SP boundary conditions.
For this purpose $c$-number solutions of the classical reflection
equation are needed. A general non-diagonal $K$ matrix satisfying
the classical reflection equation is given by (see also
\cite{ann1}) \be &&
K(\lambda) =  {\mathrm D}+{\mathrm A} + i \xi \lambda^{-1} \non\\
&& {\mathrm D} = -E_{11}-c\sum_{i=2}^{N-1}E_{ii} +E_{NN},
~~~~~{\mathrm A}=2k(E_{1N} +E_{N1}), ~~~c = 4k^2+1. \label{gsol}
\ee Apparently in the case where $k=0$ a diagonal solution is
recovered \be K(\lambda) = -\sum_{i=1}^{N-1} E_{ii}+E_{NN} + i \xi
\lambda^{-1}. \label{diag1} \ee The more general diagonal $K$
matrix is given by (see also e.g. \cite{dvg, done}) \be K(\lambda
)= -\sum_{i=1}^{l}E_{ii} + \sum_{i=l+1}^{N}E_{ii} +i\xi
\lambda^{-1}. \label{diag2} \ee The solution (\ref{diag1}) may be
seen as a special case of (\ref{diag2}) for $l=N-1$.

Henceforth we set $x=0,\ y=-L$, and we focus on the case with
diagonal boundaries provided by (\ref{diag1}), (\ref{diag2}), and
$K ={\mathbb I}$. Also $K^+(\lambda) =K(\lambda,\ \xi^+),\
K^-(\lambda)= K(-\lambda,\ \xi^-)$. The quantity under expansion
is \be && \ln \ tr \Big \{ K^+(\lambda) T(0, -L,\lambda)
K^-(\lambda) \hat T(0, -L, \lambda) \Big \} = \non\\ && \ln\ tr
\Big \{ (1+\hat W(0))^{-1} K^+(\lambda)(1+W(0)) e^{Z(0, -L)} (1+
W(-L))^{-1}K^-(\lambda)(1+\hat W(-L))e^{- \hat Z(0, -L)} \Big \}
\non\\ \label{gene} \ee where the objects with `hat' are simply
the same as before but now $\lambda \to -\lambda$. The technical
details of the relevant computations are presented in Appendix C.
\\
\\
{\bf (I)} Let us first consider the simple boundary conditions
described by (\ref{diag1}). Gathering all the information provided
by equations (\ref{fexp}), and by explicit computations concerning
the $i\lambda \to \infty$ expansion (see Appendix C, case (b) for
fore details) we conclude that the integrals of motion for the NLS
on the interval are given by: \be && I_1 =
-2 i \kappa \int_{-L}^0 dx\ \sum_{i=1}^{N-1}\psi_i(x) \bar \psi_i(x), \non\\
&& I_3 = -2 i \kappa \int_{-L}^0 dx\ \sum_{i=1}^{N-1}\Big (\kappa
|\psi_{i}(x)|^2 \sum_{j=1}^{N-1}|\psi_j(x)|^2 +\psi'_i(x) \bar
\psi'_i(x) \Big ) \non\\ && +2 i \xi^+ \kappa \sum_{i=1}^{N-1}
\psi_{i}(0) \bar \psi_i(0)-2 i \xi^-\kappa  \sum_{i=1}^{N-1}
\psi_i(-L) \bar \psi_i(-L), ~~\ldots \label{int2} \ee the quantity
$I_2$ as expected is trivial, as in the case of sine-Gordon model
on the half line. Recall that in the whole line the quantity $I_2$
corresponds essentially to the momentum, which is not a conserved
quantity any more. To obtain the number of particles and the
Hamiltonian we simply have to divide by $-2i\kappa $ \be {\cal N}=
-{I_1 \over 2i\kappa}, ~~~{\cal H}= -{I_3 \over 2i\kappa}
~~~~\mbox{and} ~~~~\{{\cal H},\ {\cal N}\}=0. \label{new} \ee It
is clear that different choices of boundary conditions lead to
distinct boundary contributions to the integrals of motion. A more
detailed description of complicated diagonal and non diagonal
boundaries is presented in Appendix C. Notice also that for $N=2$
the boundary Hamiltonian presented in \cite{sklyanin} is
recovered. Of course we could have considered Shcwartz boundary
conditions at $x=-L$ i.e. $\psi(-L),\ \bar \psi(-L) =0$ and
trivial right boundary $K^- ={\mathbb I}$ (that is the system is
considered on the half line), then the boundary terms appearing at
$x=-L$ in the expressions of the integrals of motion would
disappear.

As already mentioned the equations of motion will be derived based
on the existence of a boundary Hamiltonian rather than on the
existence of a modified Lax pair. In general, among the integrals
of motion there exists a Hamiltonian (\ref{int2}) such that the
relations below \be && {\partial \psi_i(x,t)\over
\partial t} = \Big \{{\cal H}(0,-L),\ \psi_i(x,t) \Big \}, ~~{\partial
\bar \psi_i(x,t)\over \partial t} = \Big \{{\cal H}(0,-L),\ \bar
\psi_i(x,t) \Big \}, \non\\ && -L \leq x \leq 0 \label{eqmo} \ee
give rise to the classical equations of motion. Indeed considering
the Hamiltonian ${\cal H}$ (\ref{int2}), (\ref{new}), we end up
with the following set of equations \be && i {\partial \psi_i(x,t)
\over
\partial t} = - {\partial^2 \psi_i(x,t)\over  \partial^2 x} +2\kappa
\sum_{j=1}^{N-1}|\psi_{j}(x,t)|^2\psi_i(x,t) \non\\ && \Big
({\partial \psi_{i}(x,t) \over
\partial x} -\xi^+\psi_{i}(x,t)\Big )_{x=0}=\Big ({\partial \psi_{i}(x,t)
\over \partial x} -\xi^-\psi_{i}(x,t) \Big )_{x=-L}=0, \non\\ && i
\in \{1, \ldots ,N-1\}.  \label{eqmo2} \ee In general the boundary
Hamiltonian for the generalized NLS model may be expressed as \be
{\cal H}= \int_{-L}^0 dx\ \sum_{i=1}^{N-1}\Big (\kappa
|\psi_{i}(x)|^2 \sum_{j=1}^{N-1}|\psi_j(x)|^2 +\psi'_i(x) \bar
\psi'_i(x)\Big ) +{\cal B} \ee where ${\cal B}$ is the boundary
potential. One may write the equations of motion for a generic
boundary potential ${\cal B}$. It is clear that the bulk part
remains intact as in (\ref{eqmo2}), and what is only modified is
the boundary conditions at $x=0,\ x=-L$ depending naturally on
${\cal B}$, i.e. \be \Big ({\partial \psi_{i}(x,t) \over
\partial x} +{\partial {\cal B} \over \partial \bar \psi_{i}}\Big )_{x=0}=\Big ({\partial \psi_{i}(x,t)
\over \partial x} + {\partial {\cal B} \over \partial \bar
\psi_{i}}\Big )_{x=-L}=0 \ee Two more examples of diagonal
boundaries are presented below:
\\
\\
{\bf (II)} Consider the boundary conditions described by
(\ref{diag2}). The corresponding contributions to the integrals of
motion due to the presence of non trivial boundaries are computed
in Appendix C, case (b). In this case the boundary potential (see
(\ref{last})) is given by \be {\cal B} &=& -\sum_{i=l^+ +
1}^{N-1}\Big (\psi_i(0) \bar \psi'_i(0) + \psi'_i(0) \bar
\psi_i(0)\Big ) - \xi^{+} \sum_{i=1}^{l^+} \psi_{i}(0) \bar
\psi_i(0) \non\\ &+& \sum_{i=l^- + 1}^{N-1}\Big (\psi_i(-L) \bar
\psi'_i(-L) + \psi'_i(-L) \bar \psi_i(-L) \Big ) + \xi^{-}
\sum_{i=1}^{l^-} \psi_{i}(-L) \bar \psi_i(-L), \ee and
consequently the boundary conditions for the equations of motion
at $x=0,\ x=-L$ now read as: \be && \Big ({\partial \psi_{j^+}(x)
\over \partial x} -\xi^{+}\psi_{j^+}(x) \Big )_{x=0} = \Big
({\partial \psi_{j^-}(x) \over \partial x} -\xi^{-}\psi_{j^-}(x)
\Big )_{x=-L}=0, ~~~~j^{\pm} \in \{1, \ldots, l^{\pm}\} \non\\ &&
\psi_{j^+}(0) = \psi_{j^-}(-L) =0, ~~~~~j^{\pm} \in \{l^{\pm} +1,
\ldots, N-1 \}. \ee The previous case (I) may be seen as a special
case of the more general diagonal boundary conditions by setting
$l^{\pm} =N-1$. Ultimately, one would like to investigate the SP
boundary conditions in the context of affine Toda field theories,
something that has not been achieved up to date. In this case, it
is naturally anticipated that the corresponding equations of
motion should explicitly depend on the parameters $\xi^{\pm},\
l^{\pm}$, contrary to the case analyzed in \cite{durham}, where no
extra {\it free} parameters associated to the boundaries occur. It
is also worth stressing that in the context of integrable spin
chains the integers $l^{\pm}$ appear explicitly in the
corresponding Hamiltonian as well as in the associated symmetry of
the model. More precisely, it was shown in \cite{done} that the
open spin chain with diagonal boundary conditions associated to
integers $l^{\pm} = l$ is $gl_{l} \otimes gl_{N-l}$ invariant (or
${\cal U}_q(gl_l)\otimes {\cal U}_q(gl_{N-l})$ invariant in the
trigonometric case). The symmetry breaking for the quantum $gl_N$
NLS model due to presence of non trivial integrable boundaries is
also discussed in \cite{miraso}.
\\
\\
{\bf (III)} Finally we consider the case where $K^{\pm} = {\mathbb
I}$ (Appendix C, case (c)). The boundary potential in this case is
\be {\cal B} = - \sum_{i=1}^{N-1} \Big (\psi'_i(0) \bar \psi_i(0)
+ \psi_i(0) \bar \psi'_i(0)\Big ) + \sum_{i=1}^{N-1}  \Big
(\psi_i'(-L) \bar \psi_i(-L) + \psi_i(-L) \bar \psi'_i(-L)\Big )
\ee and apparently we end up with simple Dirichlet boundary
conditions \be \psi_i(0) = \psi_i(-L) =0, ~~~i\in \{1, \dots N-1
\}. \ee Note that the $N=2$ case in particular was investigated
classically on the half line in \cite{tarasov}, whereas the NLS
equation on the interval was studied in \cite{fokas}.

\subsubsection{SNP boundary conditions}

Recall that in this case the object under consideration,
compatible with the underlying algebra, that is the classical
version of the twisted Yangian, is \be \ln\ tr \Big \{ K^+\
T(0,-L,\lambda)\ K^-\ V T^t(0,-L,-\lambda) V \Big \}\label{recal}
\ee and we choose here for simplicity $K^{\pm}={\mathbb I}$. Note
however that a generic solution of the classical twisted Yangian
is given by any matrix $K=\pm K^t$ (see \cite{ann2}). We shall
choose in what follows $V = \mbox{antid}(1, \ldots,1)$. By
expanding (\ref{recal}) in powers of $\lambda^{-1}$, along the
lines described in Appendix C, explicit expressions for the
integrals of motion are entailed (see (\ref{fexpb}), (\ref{fnl})).
It is worth pointing out, bearing in mind expressions
(\ref{fexpb}), that non-local contributions to the integrals of
motion arise. This is quite an intriguing fact and it definitely
merits further investigation, which however will be undertaken in
a forthcoming work. Nevertheless, based on the formulas
(\ref{fexpb}), (\ref{fnl}) we may explicitly express the first
non-trivial conserved quantity, which is somehow a `modified'
number of particles, i.e. \be {\cal N}_m =
\sum_{i=1}^{N-1}\int_{-L}^0 dx\ \psi_{i}(x) \bar \psi_i(x)
+\int_{-L}^0 dx\ \psi_{1}(x) \bar \psi_1(x).\ee Notice that the
SNP boundary modify dramatically the number of particles (see
(\ref{first}), (\ref{clas})). Indeed, the variation due to the
integrable boundary conditions is not limited to the addition of
certain boundary terms to the bulk quantity, as is customary, but
it gives rise to an alteration of the bulk expression itself. This
is a very interesting and definitely non-conventional aspect that
has not been encountered before, especially  in the context of
continuum integrable theories. Note finally that in the special
case $N=2$ the `modified' number of particles reduces to the usual
number of particles, which is a conserved quantity for the $sl_2$
NLS model with diagonal boundary conditions (see (\ref{int2})).

\subsection{The NLS model with reflecting impurity}

A physically relevant example will be discussed in what follows.
More precisely we shall restrict our attention to the usual NLS
model, and within the framework described in the previous section
we shall examine the problem of reflecting impurities attached to
the ends of the system. According to \cite{sklyanin} one may
consider a more general solution of the reflection equation.
Consider the classical Lax operator satisfying  \be \Big
\{{\mathbb L}_1(\lambda_1),\ {\mathbb L}_2(\lambda_2) \Big \}
=\Big [r_{12}(\lambda_1-\lambda_2),\ {\mathbb L}_1(\lambda_1)
{\mathbb L}_2(\lambda_2) \Big ], \label{clf} \ee recall
$r(\lambda)$ is given in (\ref{rr}). For example consider the
${\mathbb L}$ operator associated to the classical Lie algebra
$sl_2$: \be {\mathbb L}(\lambda) = \left(
\begin{array}{cc}
\lambda +S_3  &S_1-iS_2 \\
S_1+iS_2      &\lambda- S_3
\end{array} \right ) \label{lcl} \ee where apparently $S_a$ obey \be \{ S_a,\ S_b \} =-i
\sum_{i=1} \varepsilon_{abc}S_{c}\ee $\varepsilon$ being the usual
antisymmetric tensor. One may easily express the later matrix in
terms of canonical variables $({\mathrm x},\ {\mathrm X})$, i.e.
\be  {\mathbb L}(\lambda) = \left(
\begin{array}{cc}
\lambda+{\mathrm x}{\mathrm X}-\rho  &-{\mathrm x}^2{\mathrm X}+ 2 \rho {\mathrm x} \\
{\mathrm X }               &\lambda-{\mathrm x}{\mathrm X}+\rho
\end{array} \right ). \label{lcl2} \ee Degenerate cases of the matrix above are for
instance the Toda chain and the DST model (see e.g.
\cite{sklyaninblac} and references therein) with Lax operators
given by \be  {\mathbb L}^{Toda}(\lambda) = \left(
\begin{array}{cc}
\lambda+{\mathrm X}  &-e^{\mathrm x}\\
e^{-{\mathrm x}}      &0
\end{array} \right ), ~~~~~{\mathbb L}^{DST}(\lambda) =
\left(
\begin{array}{cc}
\lambda+{\mathrm x}{\mathrm X}  &-{\mathrm x}\\
{\mathrm X}                &-1
\end{array} \right )\ee Consider the following generating function of the integrals of motion
\be \ln tr\ \Big \{{\mathbb K}^+(\lambda) T(-L, 0 ,\lambda)
{\mathbb K}^-(\lambda) \hat T(-L, 0,\lambda)\Big \} \ee  ${\mathbb
K}$ is a  generic `dynamical' type solution of the classical
reflection equation \cite{sklyanin}, i.e. \be {\mathbb
K}^{\pm}(\lambda) = {\mathbb L}(\lambda -\Theta)\
K^{\pm}(\lambda)\ {\mathbb L}^{-1}(-\lambda -\Theta) \ee ${\mathbb
L}$ can be any solution of (\ref{clf}), $K^{\pm}$ are any
$c$-number solutions of classical reflection equation. Note that
the Poisson brackets for ${\mathbb K}$ in the classical reflection
equation are considered with respect to the canonical variables
$({\mathrm x},\ {\mathrm X})$. Here we shall deal with a simple
example, that is $K^{\pm} ={\mathbb  I}$ and $\Theta^{\pm} =0$ and
${\mathbb L}$ given by (\ref{lcl2}) (for simplicity set $\rho=0$)
then it is clear that \be {\mathbb K}^{\pm}(\lambda) = \pm \lambda
+ 2\left(
\begin{array}{cc}
{\mathrm x}^{\pm}{\mathrm X}^{\pm} &-({\mathrm x}^{\pm})^2{\mathrm X}^{\pm} \\
{\mathrm X}^{\pm}  &\-{\mathrm x}^{\pm}{\mathrm X}^{\pm}
\end{array} \right ). \ee Finally the boundary contribution to the Hamiltonian
is given by (see also Appendix B) \be I^{(b)}_3 = {1 \over 3}
h_1^3 -h_1 h_2 +h_3 +{1 \over 3} \bar h_1 ^3 -\bar h_1 \bar h_2
+\bar h_3  +2i \kappa  \psi(0) \bar \psi'(0)- 2i \kappa
\psi(-L)\bar \psi'(-L). \label{boco} \ee \be && h_0  = 1, ~~~h_1 =
2{\mathrm Z}^{+} , ~~~h_2 =2\kappa \psi(0)\bar \psi(0)
-2i\sqrt{\kappa}\Big ({\mathrm x}^{+} {\mathrm Z}^{+}
\psi(0)+({\mathrm x}^{+})^{-1}{\mathrm Z}^{+} \bar \psi(0)) \Big
), \non\\ && h_3= 2i\kappa \psi(0)' \bar \psi(0) +4\kappa {\mathrm
Z}^+\psi(0) \bar \psi(0) +2\sqrt{\kappa}\Big (({\mathrm
x}^{+})^{-1}{\mathrm Z}^{+}\bar \psi'(0)+{\mathrm x}^{+}{\mathrm
Z}^{+}\psi'(0)\Big ) \label{ab2} \ee where ${\mathrm Z}^{\pm}
={\mathrm x}^{\pm}{\mathrm X}^{\pm}$. Analogous expressions to
(\ref{ab2}) are given for $\bar h$, in particular $\bar h_n =(-)^n
h_n$: $~0\to -L,\ ({\mathrm x}^{+},\ {\mathrm X}^{+}) \to
({\mathrm x}^-,\ {\mathrm X}^-)$. Based on the latter expressions
the boundary part of the Hamiltonian may be deduced. Indeed,
bearing in mind that the boundary potential is given by ${\cal B}
= -{I^{(b)}_3 \over 2i\kappa}$ and taking into account
(\ref{boco}), (\ref{ab2}) we conclude \be {\cal B} = {\mathrm
b}({\mathrm x}^+, {\mathrm Z}^+, 0) - {\mathrm b}({\mathrm x}^-,
{\mathrm Z}^-, -L)\ee where we define \be {\mathrm b}({\mathrm x},
{\mathrm Z},x) &=& -{2 \over \sqrt{\kappa}} \Big({\mathrm x}
{\mathrm Z}^2 \psi(x) +{\mathrm x}^{-1}{\mathrm Z}^{2} \bar
\psi(x) \Big) -{1\over i\sqrt{\kappa}}\Big ({\mathrm
x}^{-1}{\mathrm Z} \bar \psi'(x) +{\mathrm x}{\mathrm Z}
\psi'(x)\Big ) \non\\ &-&\Big (\psi'(x)\bar \psi(x) + \psi(x)\bar
\psi'(x)  \Big )-{4 \over 3i\kappa} {\mathrm Z}^{3} \ee and as
expected the boundary contribution of the Hamiltonian is solely
expressed in terms of the canonical variables ${\mathrm x}^{\pm},\
{\mathrm X}^{\pm}$ as well as the boundary values of the fields
and their derivatives (see also \cite{bade} for a similar
treatment of the classical sine--Gordon model).

\section{A quantum lattice version of NLS}

\subsection{Review on periodic lattice NLS}

Let us first present the general algebraic framework associated to
the discrete quantum version of the NLS model, introduced and
studied for the periodic case in \cite{kundu, kundubas}. In the
quantum level the key object as is well known is the ${\mathbb L}$
operator satisfying: \be R_{12}(\lambda_1 -\lambda_2)\ {\mathbb
L}_{1n}(\lambda_1)\ {\mathbb L}_{2n}(\lambda_2) = {\mathbb
L}_{2n}(\lambda_2)\ {\mathbb L}_{1n}(\lambda_1)\ R_{12}(\lambda_1
-\lambda_2) \label{rtt} \ee where the $R$-matrix associated to the
$gl_N$ Yangian is \be R(\lambda) = \lambda -i \kappa {\mathbb P},
\label{rq}  \ee and obeys of course the Yang--Baxter equation
\cite{baxter, korepin}, i.e. \be R_{12}(\lambda_1 -\lambda_2)\
R_{13}(\lambda_1)\ R_{23}(\lambda_2) = R_{23}(\lambda_2)\
R_{13}(\lambda_1)\ R_{12}(\lambda_1 -\lambda_2). \label{YBE}\ee We
shall focus in this and the subsequent section in the simplest
$sl_2$ Yangian. In this case a simple solution of equation
(\ref{rtt}) is given by (on a detailed description of the
underlying algebra see e.g. \cite{kundu, kundubas}) \be {\mathbb
L}_{0n}(\lambda) =  \left(
\begin{array}{cc}
1-i\Delta \lambda+\Delta^2 \kappa \phi_n \psi_n    &-i\Delta\sqrt{\kappa} \phi_n  \\
i\Delta \sqrt{\kappa} \psi_n   &1
\end{array}. \right ) \label{ll} \ee where the $\psi_n, \phi_n$
satisfy canonical commutation relations \be\Big [ \psi_n,\ \phi_m
\Big ] ={1\over \Delta} \delta_{nm}. \ee In fact this solution may
be thought of as the quantum version of the NLS. Indeed the
classical limit of the $L$ operator (\ref{ll}) gives ${\mathbb U}$
(\ref{lax}) (for further details see \cite{kundu}). Set $\psi_n =
\int_{x_{n}}^{x_n +\Delta} dx\ \psi(x)$ then as $\Delta \to 0$ \be
{\mathbb L}(\lambda) = 1+\Delta \tilde {\mathbb U}(x, \lambda) +
{\cal O}(\Delta^2),~~~~\mbox{where}~~~~\tilde {\mathbb
U}(x,\lambda) =  \left(
\begin{array}{cc}
i\lambda    &\sqrt{\kappa} \phi(x)  \\
\sqrt{\kappa} \psi(x)   &0
\end{array} \right ) \ee
note that $\phi(x) = \bar \psi(x)$, and $\tilde {\mathbb U}$ is
equivalent to ${\mathbb U}$ (\ref{lax}) of NLS up to a gauge
transformation i.e. \be {\mathbb U} = h \tilde {\mathbb U}h^{-1}
+h_x\ h^{-1}, ~~~h=e^{-ix {\lambda\over 2} }. \ee It is more
convenient for our purposes here to use ${\mathbb L}$ with a
rescaled spectral parameter matrix. Let us multiply (\ref{ll}) by
${i\lambda \over \Delta}$ and also set $ \zeta= {i\over \lambda}$
then the rescaled ${\mathbb L}$ matrix may be written as: \be
{\mathbb L}_{0n}(\lambda)= \left(
\begin{array}{cc}
1+{{\mathbb N}_n \over \Delta} \zeta    &-i \zeta \sqrt{\kappa}  \phi_n  \\
i \zeta \sqrt{\kappa} \psi_n   &{\zeta \over \Delta}
\end{array} \right )\ee ${\mathbb N}_n = 1+\kappa \Delta^2 \phi_n \psi_n$.
For the special value $\zeta =0$ the ${\mathbb L}$ operator
reduces to a projector \be {\mathbb L}(\zeta=0) = \left(
\begin{array}{cc}
1   &0 \\
0   &0
\end{array} \right ). \label{proj} \ee Due to the fact that the algebra
(\ref{rtt}) is equipped with a coproduct one may build tensorial
representations and construct a spin chain like system with
periodic boundary conditions, by introducing the quantities \be
T_0(\lambda) =  {\mathbb L}_{0L} \ldots {\mathbb L}_{01},
~~~\mbox{and} ~~~t(\lambda) = tr_0\ T_0(\lambda) \ee  with
$T(\lambda)$ being essentially the quantum analogous of
(\ref{trans}) and apparently by virtue of (\ref{rtt}) \be \Big
[t(\lambda),\ t(\lambda') \Big ] =0. \ee By expanding $\ln
t(\zeta)$ around $\zeta =0$ we find the corresponding involutive
quantities exactly as in the classical case \cite{kundu}. It is
easy to see that due to (\ref{proj}) $t(0) =1$, then one finds
(for more details we refer the reader to \cite{kundu}) \be \ln
t(\zeta) =  \zeta \kappa C_1 + \zeta^{2} \kappa C_2 + \zeta^{3}
\kappa C_3 +\ldots, ~~~~\Big [C_n,\ C_m \Big]=0 \ee with \be &&
C_1 = {1\over \Delta \kappa} \sum_{n=1}^L {\mathbb N}_n, \non\\ &&
C_2 = \sum_{n=1}^L p_n = \sum_{n=1}^L \Big (\phi_{n+1} \psi_{n}
-{1\over 2 \kappa \Delta^2} {\mathbb N}_n^2 \Big ), \non\\ && C_3
= {1\over \Delta} \sum_{n=1}^L {h_n} =\sum_{n=1}^L \Big
(\phi_{n+1} \psi_{n-1} -({\mathbb N}_{n} + {\mathbb
N}_{n+1})\phi_{n+1} \psi_n +(3\kappa\Delta^2)^{-1} {\mathbb N}_n^3
\Big ). \ee From the latter objects one may derive lattice
versions of the classical quantities (\ref{clas}),\be && {\cal N}
= (C_1 -L )\vert_{\Delta \to 0} \to \int dx\ \phi(x) \psi(x),
\non\\ && {\cal P} =(C_2 +{L \over 2\kappa \Delta^2})\vert_{\Delta
\to 0} \to {1\over 2}\int dx\ (\phi_x \psi - \phi \psi_x) \non\\
&& {\cal H} =-C_3 +{L \over 3\kappa \Delta^2}\vert_{\Delta \to 0}
\to \int dx\ (\phi_x \psi_x +\kappa (\phi \psi)^2). \label{lat}\ee
Notice that the expressions above are symmetric to $\psi,\ \phi$
so one can set $\phi = \bar \psi$ and obtain the familiar
expressions for the NLS system (\ref{clas}). Note also that the
existence of an obvious pseudo-vacuum allows the implementation of
Bethe ansatz techniques \cite{FT, kundu}, however our aim here is
to extend such computations in the case of the integrable open
spin chain, which is discussed in the subsequent sections.

\subsection{Open lattice NLS}

We come now to the case with open boundary conditions, which is
our main concern. The underlying algebra in this case is defined
by the reflection equation \cite{cherednik} \be R_{12}(\lambda_1
-\lambda_2)K_1(\lambda_1)R_{21}(\lambda_1
+\lambda_2)K_2(\lambda_2) = K_2(\lambda_2)R_{21}(\lambda_1
+\lambda_2)K_1(\lambda_1)R_{12}(\lambda_1 -\lambda_2).
\label{re}\ee The tensorial type solutions of the reflection
equation as well known is given by \cite{sklyanin} \be {\cal
T}_0(\lambda) = T_0(\lambda)\ K_0^-(\lambda, \xi^-,c^-)\
T_0^{-1}(-\lambda) \label{opent} \ee $K^{-}(\lambda, \xi^-, k^-) $
is $c$-number solution of the reflection equation with the most
general form given by \cite{dvg1} \be \label{k} K^{\pm}(\lambda) =
\lambda \sigma^z \pm i\xi^{\pm}+2k^{\pm} \lambda (\sigma^+
+\sigma^-).\ee Note that the explicit expression of ${\mathbb
L}^{-1}(-\lambda) =\hat {\mathbb L}(\lambda)$ is given by: \be
\hat {\mathbb L}_{0n}(\lambda) =  \left(
\begin{array}{cc}
 1                              &i\Delta\sqrt{\kappa} \phi_n  \\
-i\Delta \sqrt{\kappa} \psi_n   &i\Delta \lambda+{\mathbb
N}_n-\kappa\Delta
\end{array} \right ). \label{ll-}\ee The corresponding generating function of the
conserved quantities of the open system is \be t(\lambda) = tr_0
\Big \{K_0^+(\lambda, \xi^+,k^+)\ {\cal T}_0(\lambda) \Big \} \ee
$K^+(\lambda, \xi^+,k^+) = K(-\lambda -i, \xi^+, k^+)^t$, $K$ is
again a $c$ number solution of the reflection equation. And due to
(\ref{re}) it is clear that integrability of the quantum system is
ensured i.e. \be \Big [t(\lambda),\ t(\lambda') \Big ]=0,
~~~\lambda,\ \lambda' \in {\mathbb C}.\ee

In the remaining of this section we specify the spectrum of the
lattice NLS model with diagonal boundary conditions by means of
the Bethe ansatz technique \cite{FT}. Focusing on diagonal
boundaries should be sufficient given that in \cite{ann1} the
spectral equivalence between systems with diagonal and non
diagonal boundaries was shown by means of appropriate gauge
transformations, but only for spin chains associated to the
fundamental representation of $sl_2$. Presumably there exist
suitable gauge transformations for the system under consideration,
such that the spectral equivalence is guaranteed. We shall further
comment on this point on a separate publication. When both
boundaries are diagonal there exists an obvious reference state
for the transfer state \be \Omega = \bigotimes_{n=1}^N \varpi_n
~~~~\mbox{with} ~~~\psi_n\ \varpi_{n}=0. \ee Based on this
observation one may in a straightforward manner derive the
spectrum and the corresponding Bethe ansatz equations. We provide
directly the results avoiding the technical details (for a
detailed description we refer the reader to \cite{sklyanin}). The
spectrum of the transfer matrix is given by \be  &&
\Lambda(\lambda) = {\mathrm g}(\lambda) {\mathrm
b}_1(\lambda)\prod_{j=1}^{M} {(\lambda -\lambda_{j}
+i\kappa)(\lambda +\lambda_{j})\over(\lambda -\lambda_{j}
)(\lambda +\lambda_{j}-i\kappa)} +  {\mathrm h}(\lambda) {\mathrm
b}_2(\lambda)\prod_{j=1}^{M} {(\lambda -\lambda_{j}
-i\kappa)(\lambda +\lambda_{j}-2i\kappa )\over (\lambda
-\lambda_{j} )(\lambda +\lambda_{j}-i\kappa)} \non\\ \label{spec}
\ee where we define \be &&{\mathrm g}(\lambda)=
(-i\lambda\Delta+1)^L,~~~{\mathrm h}(\lambda)=(i \lambda \Delta +
1 +\kappa \Delta)^L, \non\\ && {\mathrm b}_{1}(\lambda)= {\lambda
-i\kappa \over \lambda -{i\kappa \over 2}}(\lambda
+i\xi^-)(-\lambda+i\xi^+), ~~~{\mathrm b}_2(\lambda) ={\lambda
\over \lambda -{i\kappa \over
2}}(-\lambda+i\xi^-+i\kappa)(\lambda-i\kappa +i\xi^+)  \non\\ \ee
The corresponding Bethe ansatz equations arising as analyticity
conditions on the spectrum are: \be  {{\mathrm g} (\lambda_i
+{i\kappa \over 2}) \over {\mathrm h}(\lambda_i+{i\kappa \over
2})} {{\mathrm b}_1(\lambda_i +{i\kappa \over 2}) \over {\mathrm
b}_2(\lambda_i +{i\kappa \over 2})} = -\prod_{j=1}^M{\lambda_i
-\lambda_{j} -i\kappa \over \lambda_i-\lambda_{j}+i\kappa }\
{\lambda_i +\lambda_{j} -i\kappa \over \lambda_i+\lambda_{j}+
i\kappa }. \label{bae}\ee Notice that although we deal with an
open spin chain and one would expect a leading order of $2L$, we
see a leading order of $L$ exactly as in the periodic case. The
same phenomenon occurs in the boundary lattice Liouville model
\cite{doikou}, and is presumably associated to the degenerate
nature of the ${\mathbb L}$ matrix (see also similar comments in
\cite{sklyaninblac}). It should be emphasized that the Bethe
ansatz equations (\ref{bae}) are of particular significance given
that their thermodynamic analysis yields for instance
consequential information regarding the corresponding bulk as well
as boundary exact $S$ matrices of the model.

\section{The generalized lattice NLS}

We shall deal in what follows with the lattice quantum version of
the $gl_N$ NLS model. Recall that the $gl_N$ Yangian $R$ matrix,
solution of the Yang--Baxter equation (\ref{YBE}), is given in
(\ref{rq}). The relevant ${\mathbb L}$ operator in this case is
given by (see also \cite{kundu}) \be {\mathbb L}(\lambda) =
(-{i\lambda \over \kappa}+\sum_{j=1}^{N-1}\phi^{(j)}
\psi^{(j)})E_{11} +\sum_{j=2}^N E_{jj}
+\sum_{j=2}^{N}(\phi^{(j-1)}E_{1j}+ \psi^{(j-1)}E_{j1}).\ee Notice
that here we set implicitly $i \Delta \sqrt{\kappa} =1 $. It will
be also useful for the following to define \be \hat {\mathbb
L}(\lambda) = V_1 {\mathbb L}^{t_{1}}(-\lambda+i\kappa \rho)V_1,
~~~~~\rho = {N \over 2}\ee  we choose here $V =\mbox{antid}(1,
\ldots, 1)$, which gives rise to the following explicit form: \be
\hat {\mathbb L}(\lambda) = ({i\lambda \over \kappa} +\rho
+\sum_{j=1}^{N-1} \phi^{(j)}\psi^{(j)})E_{NN} +\sum_{j=1}^{N-1}
E_{jj} +\sum_{j=2}^{N}(\phi^{(j-1)}E_{\bar j N}+ \psi^{(j-1)} E_{N
\bar j})\ee where $\bar j = N-j+1$.  Recall that in general we
could have chosen any $V$ such that $V^2 ={\mathbb I}$. We shall
focus hereafter in the case of SNP boundary conditions, given that
they are not so widely known compared to the SP ones, especially
in the context of integrable lattice models. The main objective in
this section is to derive the exact spectrum and the corresponding
Bethe ansatz equations. Note that in the SNP case the underlying
algebra is defined by the following relation, (twisted Yangian,
see e.g. \cite{molev}) \be R_{12}(\lambda_1 -\lambda_2)\
K_{1}(\lambda_1)\ \bar R_{21}(\lambda_1+\lambda_2)\ K_2(\lambda_2)
= K_{2}(\lambda_2)\ \bar R_{12}(\lambda_1 +\lambda_2)\
K_{1}(\lambda_1)\ R_{21}(\lambda_1 -\lambda_2)\label{ty} \ee and
in analogy to the classical case we define \be \bar R(\lambda) =
V_1\ R^{t_1}_{12}(-\lambda +i\rho \kappa )\ V_1.\ee The generating
function of the quantum integrals of motion in this case is
defined as: \be && t(\lambda)= tr\ \Big \{ K^{+}(\lambda)\
T(\lambda)\ K^-(\lambda)\ \hat T(\lambda)\Big \}, ~~~~\mbox{with}
\non\\ && T(\lambda)={\mathbb L}_{0L}(\lambda) \ldots {\mathbb
L}_{01}(\lambda), ~~~~\hat T(\lambda)= \hat {\mathbb
L}_{01}(\lambda) \ldots \hat {\mathbb L}_{0L}(\lambda)\label{t2}
\ee $K^{\pm}$ are $c$-number solutions of the twisted Yangian
(\ref{ty}). In fact, it was shown in \cite{ann2} that any matrix
$K =\pm K^t$ is a solution of the twisted Yangian. In Appendix D
an explicit computation of local integrals of motion for
particular boundary conditions is presented. Based on these
findings we present the explicit form of the boundary momentum in
the case where $K^{\pm} =V$, i.e. \be {\cal P}_d= -{i\kappa \over
2} \Big (\sum_{n=1}^L {\mathbb N}_n^{2} -
2\sum_{n=1}^{L-1}\sum_{j=1}^{N-1} \psi_n^{(j)} \phi_{n+1}^{(j)} +
\sum_{j=1}^{N-1}
(\psi_L^{(j)}\psi_L^{(j)}+\phi_1^{(j)}\psi_1^{(j)})  \Big ).
\label{pp2} \ee We could have of course considered $K^{\pm}
\propto {\mathbb I}$, which is the case will be examined
subsequently, but for the sake of simplicity we considered the
aforementioned boundary conditions, which from a technical point
of view are much easier to deal with. By comparing the bulk
momentum given in Appendix D (\ref{p1}) with (\ref{pp2}) we
conclude that the periodic terms in (\ref{p1}) are replaced
essentially by the last two boundary terms in (\ref{p2}), whereas
the bulk part remains intact. Following the logic of (\ref{lat}),
it is expected that the expression (\ref{pp2}) should be a
regularization of the continuum generalized NLS model momentum
with particular SNP boundary conditions, exactly as it happens for
the periodic NLS model (indeed compare the bulk continuum
expression (\ref{first}) with the discrete periodic analogous
(\ref{bulkd})). Comments on higher conserved charges may be also
found in Appendix D.

To deduce the spectrum and Bethe ansatz equations of the
generalized NLS model with SNP boundary conditions we shall
restrict our attention to another simple case i.e.
$K^{\pm}={\mathbb I}$. Again the spectral equivalence between
systems with diagonal and non-diagonal boundaries is discussed in
\cite{ann2, crdo}, for spin chains associated to the fundamental
representation of $gl_N$. The first step toward the
diagonalization of the transfer matrix (\ref{t2}) is the
derivation of a reference state. Indeed, in this case there exists
an obvious reference state, that is \be \Omega =
\bigotimes_{n=1}^L \varpi_{n}: ~~~\psi_n^{(i)}\ \varpi_{n} = 0,
~~~~~n \in \{1, \ldots, L\}, ~~~~~i \in \{1, \ldots N-1 \}. \ee
The corresponding eigenvalue may be easily derived using the fact
that ${\mathbb L},\ \hat {\mathbb L}$ and consequently $T,\ \hat
T$ satisfy \be \hat T_{1}(\lambda_1)\ \bar
R_{12}(\lambda_{1}+\lambda_2)\ T_{2}(\lambda_2) = T_2(\lambda_2)\
\bar R_{12}(\lambda_1+\lambda_2)\ \hat T_1(\lambda_1).
\label{newa} \ee Taking into account the latter relation we
conclude that the actions of the transfer matrix on the pseudo
vacuum provides the following eigenvalue: \be
\Lambda^{(0)}(\lambda)= a^L(\lambda)\ g_1(\lambda) +
\sum_{n=2}^{N-1}g_{n}(\lambda) + \bar b^L(\lambda)\ g_{N}(\lambda)
\ee where we define \be && a(\lambda) = -{i\lambda \over \kappa},
~~~~~\bar b(\lambda) = {i\lambda \over \kappa} +\rho \non\\ &&
g_{n}(\lambda) = {\lambda -{i\kappa \over 2}(\rho -1)\over \lambda
-{i\kappa \over 2}}, ~~~1 \leq n < {N+1\over 2} \non\\ && g_{{N+1
\over 2}} =1, ~~~N ~~\mbox{odd} \non\\ &&g_{l} = g_{\bar
l}(-\lambda +i\kappa \rho), ~~~~{N+1 \over 2}< l \leq N. \ee The
functions $g_{n}$ are essentially `boundary' contributions to the
spectrum.

To determine the general eigenvalue form we shall adopt the
analytical Bethe ansatz formulation \cite{analyt}. The basic
assumption within this framework is that the structure of any
eigenvalue is similar to the pseudo-vacuum eigenvalue i.e. \be
\Lambda(\lambda)= a^L(\lambda)\ g_1(\lambda)\ A_{1}(\lambda) +
\sum_{n=2}^{N-1}g_{n}(\lambda)\ A_{n}(\lambda) + \bar
b^L(\lambda)\ g_{N}(\lambda)\ A_{N}(\lambda). \ee The so called
dressing functions $A_n$ may be explicitly determined by imposing
certain physical and algebraic requirements, such as analyticity,
crossing, etc. We do not give all the details of the formulation,
(for a more detailed description of the process we refer the
reader to \cite{menean, ann1, ann2}), but the explicit expressions
for the dressing functions are given by: \be && A_{1}(\lambda) =
\prod_{j=1}^{M^{(1)}} {\lambda +\lambda_{j}^{(1)} +{i\kappa \over
2} \over \lambda +\lambda_{j}^{(1)} -{i\kappa\over 2}  }\ {\lambda
-\lambda_{j}^{(1)} +{i\kappa\over 2} \over \lambda -
\lambda_{j}^{(1)} -{i\kappa\over 2} } \non\\ && A_{l+1} =
\prod_{j=1}^{M^{(l}} {\lambda +\lambda_{j}^{(l)} - {il\kappa\over
2} -i\kappa \over \lambda +\lambda_{j}^{(l)} -{il\kappa\over 2}}\
{\lambda -\lambda_{j}^{(l)} -{il\kappa\over 2} -i\kappa \over
\lambda - \lambda_{j}^{(l)} -{il\kappa\over 2} }\ \non\\ &&
\prod_{j=1}^{M^{(l+1)}} {\lambda +\lambda_{j}^{(l+1)}
-{il\kappa\over 2} +{i\kappa\over 2} \over \lambda
+\lambda_{j}^{(l+1)} -{il\kappa\over 2} -{i\kappa\over 2} }\
{\lambda -\lambda_{j}^{(l+1)} -{il\kappa\over 2} +{i\kappa\over 2}
\over \lambda - \lambda_{j}^{(l+1)} -{il\kappa\over
2}-{i\kappa\over 2} },~~~~~1\leq l < {N-1 \over 2} \non\\ &&
A_{l}(\lambda) =A_{\bar l}(-\lambda +i\kappa \rho), ~~~~{N-1\over
2}<l \leq N \ee  and in particular for $N=2n+1$ \be
A_{n+1}(\lambda) &= & \prod_{j=1}^{M^{(n)}} {\lambda
+\lambda_{j}^{(n)} -{in\kappa\over 2} -i\kappa \over \lambda
+\lambda_{j}^{(n)} -{in\kappa\over 2}  }\ {\lambda
-\lambda_{j}^{(n)} -{in\kappa\over 2} -i\kappa \over \lambda -
\lambda_{j}^{(n)} -{in\kappa\over 2} } \non\\ &\times&
\prod_{j=1}^{M^{(n+1)}} {\lambda +\lambda_{j}^{(n+1)}
-{in\kappa\over 2} +{i\kappa\over 2} \over \lambda
+\lambda_{j}^{(n+1)} -{in\kappa\over 2} -{i\kappa \over 2} }\
{\lambda -\lambda_{j}^{(n+1)} -{in\kappa\over 2} +{i\kappa\over 2}
\over \lambda - \lambda_{j}^{(n+1)} -{in\kappa\over
2}-{i\kappa\over 2}}.\ee Finally Bethe ansatz equations follow as
analyticity requirements upon the spectrum, and they are written
explicitly as:
\\
\\
{\bf (i) N=2n+1}: \be && \Big (a(\lambda_{i}^{(1)}+{i\kappa \over
2}) \Big )^L\delta_{l1} + (1-\delta_{l1})= -\prod_{j=1}^{M^{(l)}}
e_{2}(\lambda_{i}^{(l)} -\lambda_{j}^{(l}) e_{2}(\lambda_{i}^{(l)}
+\lambda_{j}^{(l}) \non\\ && \times
\prod_{j=1}^{M^{(l+\tau))}}e_{-1}(\lambda_{i}^{(l)}
-\lambda_{j}^{(l+\tau)}) e_{-1}(\lambda_{i}^{(l)}
+\lambda_{j}^{(l+\tau)}) \non\\ && l =1, \ldots, n-1 \non\\ &&
e_{-{1\over 2}}(\lambda_{i}^{(n)})=-\prod_{j=1}^{M^{(n)}}
e_{2}(\lambda_{i}^{(n)} -\lambda_{j}^{(n)})
e_{2}(\lambda_{i}^{(n)} +\lambda_{j}^{(n})
e_{-1}(\lambda_{i}^{(n)} -\lambda_{j}^{(n)})
e_{-1}(\lambda_{i}^{(n)} +\lambda_{j}^{(n)}) \non\\ && \times
\prod_{j=1}^{M^{(n-1)}}e_{-1}(\lambda_{i}^{(n)}
-\lambda_{j}^{(n-1)}) e_{-1}(\lambda_{i}^{(n)}
+\lambda_{j}^{(n-1)}) \label{bae2} \ee with $\tau = \pm 1$,
$~M^{(N+1)} =0$ and define $e_n(\lambda) = {\lambda -{in \kappa
\over 2} \over \lambda +{in \kappa \over 2}}$.
\\
\\
{\bf (ii) N=2n}: In this case the Bethe ansatz equations for $l =
1, \ldots n-1$ are the same as in the previous case. What is only
modified is the last set of equations, which takes the form: \be
e_{-{1\over 2}}(\lambda_{i}^{(n)})= -\prod_{j=1}^{M^{(n)}}
e_{2}(\lambda_{i}^{(n)} -\lambda_{j}^{(n)})
e_{2}(\lambda_{i}^{(n)} +\lambda_{j}^{(n)})
\prod_{j=1}^{M^{(n-1)}}e^2_{-1}(\lambda_{i}^{(n)}
-\lambda_{j}^{(n-1)}) e^2_{-1}(\lambda_{i}^{(n)}
+\lambda_{j}^{(n-1)}).\non\\ \ee Such type of boundary conditions
were first introduced in \cite{doikousnp} for the $sl_3$ spin
chain, whereas generalizations investigated in \cite{ann1, ann2}
from the physical (Bethe ansatz) as well as the algebraic point of
view. The associated symmetries were studied in detail in
\cite{ann2, doikouy, crdo}, while in \cite{doikouy} both SP and
SNP boundary conditions were examined in parallel. The interesting
observation is that the RHS of the equations above in the case
where $N=2n+1$ coincide with the ones associated to the
$osp(1|2n)$ algebra. In any case the Bethe ansatz equations
(\ref{bae2}) are somehow `folded' compared to the usual $gl_N$
ones. This is expected given that folding occurs at the algebraic
level as well (Dynkin diagrams), and only the subalgebra invariant
under charge conjugation survives after the implementation of
these rather unconventional boundary conditions (see also relevant
comments in \cite{ann2}). The case of SP boundaries can be also
treated along the same lines, and the entailed spectrum and Bethe
ansatz equations will have the usual $gl_N$ structure (the
expressions are omitted here for brevity). More precisely the LHS
of the Bethe ansatz equations will have exactly the same form as
the usual $gl_N$ BAE for an open chain (see e.g. \cite{dvg, done,
ann1}), while the RHS will depend explicitly on the actions of the
diagonal entries of the ${\mathbb L},\ {\mathbb L}^{-1}$ on the
pseudo-vacuum.

\section{Discussion}

To summarize, SP and SNP boundary conditions were studied for the
classical generalized NLS model, and the boundary integrals of
motion as well as the relevant classical equations of motion were
explicitly derived. This was the first time, to our knowledge,
that SNP boundaries were implemented in the context of the
generalized NLS model. Nevertheless, there are still several open
questions especially regarding the locality of some of the
integrals of motion for particular choices of left/right
boundaries, which however will be left for future investigations.
In the same spirit the usual ($sl_2$) NLS model with reflecting
impurities was also analyzed. Moreover, a suitable lattice version
of the generalized NLS model was considered and the SNP boundary
conditions were implemented, given that in general they are much
less studied in this context. For this choice of boundary
conditions we were able to specify the first non-trivial local
integral of motion i.e. the `boundary momentum'. We also derived
the spectrum and Bethe Ansatz equations for the simplest
left/right boundaries.

Although SP boundary conditions are somehow the obvious ones in
the framework of lattice integrable models, they have not been
really considered up to now in classical continuum integrable
theories. Therefore, it will be our next goal to impose the SP
boundary conditions to other well known classical systems such as
(massless) affine Toda field theories, principal chiral models,
etc. In addition, the investigation of the generalized (m)KdV
hierarchies \cite{DS} with integrable boundaries is another very
interesting direction to pursue together with their quantization
into an appropriate lattice version (see e.g. \cite{fior} and
references therein). Once this point is clarified, the study of
the underlying dynamical symmetries constrained by integrable
boundary conditions could be discussed in full generality along
the lines described in \cite{fiost}, and this would definitively
shed new light on the character of the different integrable
boundary conditions. More precisely, it would be of great
consequence to examine how the so called hidden symmetries
constructed in \cite{fiost} are modified in the presence of
non-trivial integrable boundaries, and in particular in the case
of (quantum) twisted Yangians. Finally, an interesting direction
to pursue is the explicit derivation of the modified Lax pairs by
means of the `boundary' integrals of motion. We hope to report on
all these issues in forthcoming publications.
\\
\\
\noindent{\bf Acknowledgments:} This work was supported by
national and local (Bologna section) INFN through grant TO12,
Italian Ministry of University and Research through a PRIN grant,
NATO Collaborative Linkage Grant PST.CLG.980424,  and the European
Network `EUCLID'; `Integrable models and applications: from
strings to condensed matter', contract number
HPRN--CT--2002--00325.

\appendix

\section{Appendix}

In this appendix the classical integrability for models with both
types of boundary conditions, SP and SNP, is reviewed. The first
step in order to prove the classical integrability is to show that
the quantities introduced in (\ref{reps}) are indeed
representations of the algebras defined by (\ref{refc}). To
achieve this we shall need in addition to (\ref{basic}) the
following set of algebraic relations emerging essentially from
(\ref{basic}), i.e. \be && \Big \{ \hat T_1(\lambda_1),\ \hat
T_2(\lambda_2) \Big \} = r_{12}(\lambda_1 -\lambda_2)\ \hat
T_1(\lambda_1)\ \hat T_2(\lambda_2) - T_1(\lambda_1)\ \hat
T_2(\lambda_2)\ r_{12}(\lambda_1 -\lambda_2) \non\\ && \Big \{
T_1(\lambda_1),\ \hat T_2(\lambda_2) \Big \} =T_1(\lambda_1)\
r_{12}^*(\lambda_1 +\lambda_2)\ \hat T_2(\lambda_2) - \hat
T_2(\lambda_2)\ r_{12}^*(\lambda_1+\lambda_2) \  T_1(\lambda_1)
\non\\ && \Big \{ \hat T_1(\lambda_1),\  T_2(\lambda_2) \Big \} =
\hat T_1(\lambda_1)\ r_{12}^*(\lambda_1 +\lambda_2)\
T_2(\lambda_2) - T_2(\lambda_2)\ r_{12}^*(\lambda_1+\lambda_2) \
\hat T_1(\lambda_1) \label{algebr}  \ee Our aim now is to show
that (\ref{refc}) are satisfied by (\ref{rep2}): \be  && \Big \{
{\cal T}_1(\lambda_1),\ {\cal T}_2(\lambda_2) \Big \} = \Big \{
T_1(\lambda_1) K^{-}_1(\lambda_1) \hat T_1(\lambda_1), \
T_2(\lambda_2) K^{-}_2(\lambda_2) \hat T_2(\lambda_2) \Big \} =
\ldots  \non\\  && = T_1(\lambda_1) T_2(\lambda_2) \Big (
-K_1(\lambda_1) K_2(\lambda_2) r_{12}(\lambda_1 -\lambda_2)
+r_{12}(\lambda_1 -\lambda_2)K_1(\lambda_1) K_2(\lambda_2) \non\\
&& + K_1(\lambda_1)r^*_{12}(\lambda_1 +\lambda_2) K_2(\lambda_2)
-K_2(\lambda_2) r_{12}^*(\lambda_2) K_1(\lambda_1) \Big )\hat
T_1(\lambda_1) \hat T_2(\lambda_2) \non\\ && + {\cal
T}_1(\lambda_1) r^*_{12}(\lambda_1 +\lambda_2) {\cal
T}_2(\lambda_2) - {\cal T}_2(\lambda_2) r^*_{12}(\lambda_1
+\lambda_2){\cal T}_1(\lambda_1) \non\\ && r_{12}(\lambda_1
-\lambda_2){\cal T}_1(\lambda_1){\cal T}_2(\lambda_2) -{\cal T}_1
(\lambda_1){\cal T}_2(\lambda_2) r_{12}(\lambda_1 -\lambda_2)
\label{fin} \ee and making use of (\ref{kso}), (\ref{cno}) and
(\ref{algebr}) we end up to (\ref{refc}). Recall also that
$c$-number solutions of the above equations satisfy the following
\be \Big [ r_{12}(\lambda_1-\lambda_2),\
K_{1}(\lambda_1)K_2(\lambda_2)\Big ] = K_{2}(\lambda_2)
r^*_{12}(\lambda_1+\lambda_2)K_1(\lambda_1)-K_{1}(\lambda_1)
r^*_{12}(\lambda_1+\lambda_2)K_2(\lambda_2), \non\\ \label{cno}
\ee which is equivalent to (\ref{kso}).

We may now show exploiting (\ref{refc}) and (\ref{cno}) the
classical integrability (\ref{bint}). Indeed consider the
following  object \be  \Big \{ K^+_1(\lambda_1) {\cal
T}_1(\lambda_1),\ K_2^+(\lambda_2) {\cal T}_2(\lambda_2) \Big \}
\ee then taking the trace in both spaces 1 and 2, and considering
the defining relations (\ref{refc}), (\ref{kso}) we end up with
\be \Big \{t(\lambda_1),\ t(\lambda_2) \Big \}& =& tr_{12}\Big (
K_1^+(\lambda_1)K_2^+(\lambda_2)r_{12}(\lambda_1-\lambda_2) {\cal
T}_1(\lambda_1) {\cal T}_2(\lambda_2) \non\\ &-&
K_1^+(\lambda_1)K_2^+(\lambda_2){\cal T}_1(\lambda_1) {\cal
T}_2(\lambda_2) r_{12}(\lambda_1-\lambda_2) \non\\ &+&
K_1^+(\lambda_1)K_2^+(\lambda_2){\cal T}_1(\lambda_1)
r^*_{12}(\lambda_1+\lambda_2){\cal T}_2(\lambda_2) \non\\ &-&
K_1^+(\lambda_1)K_2^+(\lambda_2){\cal T}_2(\lambda_2)
r^*_{12}(\lambda_1+\lambda_2){\cal T}_1(\lambda_1) \Big ). \ee
Finally moving appropriately the factors of the products within
the trace and using (\ref{cno}) it is straightforward to show \be
\Big \{t(\lambda_1),\  t(\lambda_2) \Big \}=0, ~~~~\lambda_1,\
\lambda_2 \in {\mathbb C} \ee and this concludes our proof.
Similar arguments hold also in the case of dynamical boundaries
discussed in section 3.3.

\section{Appendix}

We present here some technical details on the derivation of the
conserved quantities for the generalized NLS model on the full
line. The first step is to insert the ansatz (\ref{exp0}) in
equation (\ref{dif1}).  Then we separate the diagonal and off
diagonal part and obtain the following expressions: \be && Z' =
\lambda {\mathbb U}_1 +({\mathbb U}_{0}W)^{(D)} \non\\ && W' +WZ'
= {\mathbb U}_{0} +({\mathbb U}_0 W)^{(O)} +\lambda{\mathbb U}_1 W
\label{DO} \ee where the superscripts $(D),\ (O)$ denote the
diagonal and off diagonal part of the product ${\mathbb U}_0W$.
Recall that $W = \sum_{i\neq j} W_{ij} E_{ij}, ~~Z= \sum_{i}Z_{ii}
E_{ii}$ then it is straightforward to obtain: \be && ({\mathbb
U}_0W)^{(D)} = \sqrt{\kappa} \sum_{i=1}^{N-1}\Big ( \bar \psi_i
W_{Ni} E_{ii} + \psi_i W_{iN} E_{NN} \Big ) \non\\ && ({\mathbb
U}_0W)^{(O)} = \sqrt{\kappa} \sum_{i\neq j,\ i\neq N,\ j\neq N}
\Big (\bar \psi_i W_{Nj} E_{ij} + \psi_{i} W_{ij}E_{ Nj} \Big ).
\label{contr} \ee Substituting the latter expressions
(\ref{contr}) in (\ref{DO}), we obtain \be Z(L, -L, \lambda) =
-i\lambda L \Big (\sum_{i=1}^{N-1}E_{ii} -E_{NN} \Big )
+\sqrt{\kappa} \sum_{i=1}^{N-1}\int_{-L}^{L}\ dx \Big (\bar \psi_i
W_{Ni} E_{ii} +\psi_i W_{iN} E_{NN}\Big ) \ee And recalling that
the leading contribution in the expansion of ($\ln\ tr T$), --$T$
is given in (\ref{exp0})-- as $i\lambda \to \infty $ is coming
from the $Z_{NN}$ term we conclude: \be Z_{NN}(L, -L, \lambda) =
i\lambda L + \sqrt{\kappa} \sum_{i=1}^{N-1}\int _{-L}^{L} dx\
\psi_{i}(x) W_{iN}(x) \label{Z}\ee Due to (\ref{Z}) it is obvious
that in this case it is sufficient to derive the coefficients
$W_{iN}$ only. In any case one can show that the coefficients
$W_{ij}$ satisfy the following equations: \be  && \sum_{ i\neq j}
W_{ij}'E_{ij} -i\lambda \sum_{i\neq N} \Big (W_{Ni} E_{Ni}
-W_{iN}E_{iN}\Big ) +\sqrt{\kappa}\sum_{i\neq N}\Big (\bar
\psi_{i} W_{Ni}^2E_{Ni} + \psi_i W_{iN}^2 E_{iN} \Big ) =\non\\ &&
\sqrt{\kappa}\sum_{i\neq N}\Big (\bar \psi_{i} E_{iN} + \psi_{i}
E_{Ni}\Big ) + \sqrt{\kappa} \sum_{i\neq j,\ i\neq N,\ j\neq
N}\Big (\bar \psi_i W_{Nj} E_{ij} +\psi_i W_{ij}E_{Nj} \Big )
\non\\ && - \sqrt{\kappa} \sum_{i\neq j,\ i\neq N,\ j\neq N}\Big
(\bar \psi_{j}W_{Nj}W_{ij} E_{ij} +\psi_i W_{iN} W_{jN} E_{jN}\Big
) \label{rec1} \ee Finally setting $W_{ij} =\sum_{n=1}^{\infty}
{W_{ij}^{(n)} \over \lambda^n}$ and using (\ref{rec1}) we find
expressions for $W_{iN}^{(n)}$ reported in (\ref{ref2}). In the
case with integrable boundary conditions, we shall need in
addition to (\ref{ref2}) the following objects: \be &&
W_{Ni}^{(1)} = i\sqrt{\kappa}\psi_i, ~~~~W_{Ni}^{(2)} = -i
W^{'(1)}_{Ni} +\sum_{i\neq j,\ i\neq N,\ j\neq N}W^{(1)}_{N
j}W_{ji}^{(1)},~~~~W_{ji}^{'(1)} = iW_{jN}^{(1)} W_{Ni}^{(1)}
\non\\ && W_{Ni}^{(3)} = - iW^{'(2)}_{Ni}
+W_{iN}^{(1)}W_{Ni}^{(1)}W_{Ni}^{(1)} + \sum_{i\neq j,\ i \neq N,
j\neq N} W_{Nj}^{(1)}W_{ji}^{(2)} \non\\ && W_{ij}^{'(2)} =
iW_{iN}^{(1)} W_{Nj}^{(2)}-i W_{jN}^{(1)} W_{Nj}^{(1)}
W_{ij}^{(1)}. \label{reff}\ee

\section{Appendix}

In what follows we evaluate the boundary terms contributing to the
Hamiltonian for right and left boundary described by the more
general, diagonal and non-diagonal, solutions of the reflection
equation (SP boundary conditions). Moreover, for the SNP boundary
conditions we identify the corresponding integrals of motions, and
we explicitly evaluate the first non-trivial charge.

\subsection{SP boundary conditions}

We shall expand the generic object (\ref{gene}) keeping of course
only diagonal contributions. More precisely, as in the bulk case
due to the fact that the leading order is $e^{i\lambda L}$ as
$i\lambda \to \infty$ the only non negligible part is coming from
the $E_{NN}$ terms, hence we shall only consider such
contributions: \be && \Big [(1+\hat
W(0,\lambda))^{-1}K^+(\lambda)(1+W(0,\lambda))\Big ]_{NN} =
\sum_{n=0}^{\infty}{h_{n} \over \lambda^{n}} , \non\\  && \Big
[(1+ W(-L,\lambda))^{-1}K^-(\lambda)(1+\hat W(-L,\lambda))\Big
]_{NN} = \sum_{n=0}^{\infty}{\bar h_{n} \over \lambda^{n}}  \non\\
&& \Big [Z(0,-L,\lambda)- \hat Z(0,-L,\lambda)\Big ]_{NN} = i
\lambda L + \sum_{n=1}^{\infty} (1-(-)^{n}){Z_{NN}^{(n)}(0,-L)
\over \lambda^n}\label{expa2} \ee Again considering only the
contribution of the term $e^{i\lambda L}$ as $i\lambda \to \infty$
we end up with the following expression \be && \ln\ tr \Big
\{K^+(\lambda) T(0,-L ,\lambda) K^-(\lambda)  \hat T( 0,
-L,\lambda) \Big \}= \non\\ && i\lambda L + \sum_{n=1}^{\infty}
(1-(-)^n){Z^{(n)}_{NN}(0,-L) \over \lambda^n} +\ln \Big (
\sum_{n=0}^{\infty}{h_n +\bar h_n \over \lambda^n} +\sum_{n,m
=0}^{\infty} {h_n \bar h_m \over \lambda^{n+m}} \Big )
\label{fexp} \ee Recall from (\ref{fexp}) that the boundary
contribution lies basically in the logarithmic function, hence one
has to expand the log i.e. \be \ln (\sum_{n=0}^{\infty} {h_n +\bar
h_n \over \lambda^n}+\sum_{n,m =0}^{\infty} {h_n \bar h_m \over
\lambda^{n+m}}) = \sum_{n=0}^{\infty}{f_n \over \lambda^n}
\label{c2} \ee where $f_n$, provide essentially the boundary
contribution to the integrals of motion (plus possible total
derivatives from the bulk part) for the left right boundary
respectively. The interesting observation is that the boundary
contribution decouples nicely to terms associated to left and
right boundary separately, i.e. no mixing occurs \be && f_1 = h_1
+\bar h_1, ~~~f_2 = -{1\over 2}h_1^{2} + h_2-{1\over 2}\bar
h_1^{2} + \bar h_2 , \non\\ && f_3= {1 \over 3} h_1^3 -h_1 h_2
+h_3  +{1 \over 3} \bar h_1^3 -\bar h_1 \bar h_2 +\bar h_3 ,
~~\ldots \label{bcon} \ee
\\
\\
{\bf (a)} We first consider generic non diagonal boundary
conditions described by (\ref{gsol}). One can explicitly evaluate
$h_n$ for the generic case:
\be && h_0  = 1, ~~~h_{1} = -2ik^+\sqrt{\kappa}\Big (\bar \psi_1(0) - \psi_1(0)\Big ) +i \xi^+ , \non\\
&& h_2 =2 k^+\sqrt{\kappa}  \Big (\bar \psi_1'(0) - \psi_1'(0)\Big
)-4k^{+2} \kappa \sum_{i=2}^{N-1}
\psi_i(0) \bar \psi_i(0), \non\\
&& h_3 = 2ik^+\sqrt{\kappa} \Big (\bar \psi_1''(0)
-\psi_1''(0)\Big ) +2\kappa i \xi^+ \sum_{j=1}^{N-1} \psi_j(0)
\bar \psi_j(0)-2 i \kappa \sum_{j=1}^{N-1} \psi_j(0) \bar
\psi_j'(0) \non\\ && + 2ik^+ \kappa^{{3\over 2}}
\sum_j|\psi_j(0)|^2 \Big (\psi_1(0) - 2\bar \psi_1(0)\Big
)+2ik^+\kappa^{{3\over 2}} \bar \psi_1 \psi_1^2 -4 i\kappa k^{+2}
\sum_{j-2}^{N-1}\Big (\psi_j(0) \bar \psi_j'(0) + \psi_j'(0) \bar
\psi_j(0) \Big )  \non\\ && \ldots \label{ab} \ee In fact it is
clear from the expansion of the left and right boundary
contribution that $\bar h_n =(-1)^n h_n:\ 0 \to -L,\ \xi^+\to
\xi^-,\ k^+\to k^- $.

Given that there is an overall derivative from the bulk part of
$Z_3$, giving rise to a boundary term ($2i \kappa \sum_i \psi_i
\bar \psi_i' $), we conclude that the boundary contribution to the
conserved quantity $I_3$ (i.e. the Hamiltonian) is given by \be
I_3 = f_3 +2i \kappa \sum_{i=1}^{N-1} \bar \psi'_i(0) \psi_i(0)-2i
\kappa \sum_{i=1}^{N-1} \bar \psi_i'(-L) \psi_i(-L)=-2i\kappa
{\cal B}, \label{last} \ee where ${\cal B}$ is the boundary
potential, and recall that $f_3$ is given by (\ref{bcon}). For
diagonal boundary, terms proportional to $k^{\pm}$ apparently
disappear, and the Hamiltonian reduces to (\ref{int2}). For a
purely antidiagonal boundary only terms proportional to $k^{\pm}$
survive, all the other terms disappear.
\\
\\
{\bf (b)} The more general diagonal boundary conditions are
described by (\ref{diag2}). We associate the integers $l^{\pm}$
and the free parameters $\xi^{\pm}$ to the right/left boundaries.
In this case from the expansions (\ref{expa2}) and taking into
account (\ref{ref2}), (\ref{reff}) we find: \be && h_0=1, ~~~~h_1
=i \xi^{+}, ~~~~h_2 = 2\kappa \sum_{i=l^+ + 1}^{N-1} \psi_i(0)
\bar \psi_i(0) \non\\ && h_3= -2i\kappa \sum_{i=1}^{l^{+}}
\psi_i(0) \bar \psi'_i(0) + 2i\kappa \sum_{i=l^+ +1}^{N-1}
\psi'_i(0) \bar \psi_i(0) +2i\kappa \xi \sum_{i =1}^{N-1}
\psi_i(0) \bar \psi_i(0), ~~\ldots \ee similar expressions of
course hold for $\bar h_n$, i.e. $\bar h_n =(-1)^n h_n$:  $0\to
-L,\ l^+ \to l^-,\ \xi^+ \to \xi^-$. Taking into account
(\ref{bcon}), (\ref{last}) and derivative contribution from the
bulk ($2i\kappa \sum_{i=1}^{N-1} \psi_i \bar \psi'_i$) we conclude
that the boundary contribution to the Hamiltonian is given by: \be
I_{3} &=& 2 i \kappa \Big [\sum_{i=l^+ + 1}^{N-1}\Big (\psi_i(0)
\bar \psi'_i(0) + \psi'_i(0) \bar \psi_i(0)\Big ) -\sum_{i=l^- +
1}^{N-1}\Big (\psi_i(-L) \bar \psi'_i(-L) + \psi'_i(-L) \bar
\psi_i(-L) \Big )\Big] \non\\ &+& 2i\kappa \Big [\xi^{+}
\sum_{i=1}^{l^+} \psi_{i}(0) \bar \psi_i(0) -\xi^{-}
\sum_{i=1}^{l^-} \psi_{i}(-L) \bar \psi_i(-L)\Big ]. \ee Note that
when $l^{\pm} =N-1$ one recovers the diagonal limit of the more
general case (a).
\\
\\
{\bf (c)} Finally the case where $K^{\pm} ={\mathbb I}$ may be
treated in the same spirit. In this case we find that \be h_0=1,
~~~h_1 =0, ~~~h_2  = 2\kappa \sum_{i=1}^{N-1} \psi_i(0) \bar
\psi_i(0), ~~~h_3 =2 i \kappa \sum_{i=1}^{N-1} \psi_i'(0) \bar
\psi_i(0), ~~\ldots \ee and the corresponding boundary
contribution to the Hamiltonian is given by \be I_3 = 2 i \kappa
\Big [ \sum_{i=1}^{N-1} \Big (\psi'_i(0) \bar \psi_i(0) +
\psi_i(0) \bar \psi'_i(0)\Big ) -  \sum_{i=1}^{N-1}\Big
(\psi_i'(-L) \bar \psi_i(-L) +\psi_i(-L) \bar \psi'_i(-L)\Big
)\Big ]. \ee

\subsection{SNP boundary conditions}

Recall that in this case the object under consideration is given
in (\ref{recal}), also we consider for simplicity $K^{\pm} =
{\mathbb I}$ and we choose $V =\mbox{antid}(1,\ldots ,1)$. Before
we proceed with the evaluation of the integrals of motion let us
first introduce some useful notation \be  && \hat W_{ij}(\lambda)
= W_{\bar j \bar i}(-\lambda),
~~~~\hat Z_{ii}(\lambda) =Z_{\bar i \bar i}(-\lambda), \non\\
&& \mbox{and} ~~~(1+W(\lambda))^{-1} = 1 + F(\lambda)
~~~\mbox{where} ~~~F(\lambda) = \sum_{n=1}^{\infty} {{\mathrm
f}^{(n)} \over \lambda^n}, \label{nots0} \ee recall $\bar j =
N-j+1$, and also one may easily compute that \be {\mathrm f}^{(1)}
= -W^{(1)}, ~~~~{\mathrm f}^{(2)} = (W^{(1)})^2 -W^{(2)},
~~~\ldots \label{nots} \ee We shall need the following
contributions in order to evaluate the corresponding integrals of
motion: \be && (1+\hat W(0,\lambda))(1+W(0,\lambda)) =
1+H(\lambda), \non\\  && (1+ W(-L,\lambda))^{-1}(1+\hat
W(-L,\lambda))^{-1} = 1+\bar H(\lambda).  \label{expa2b} \ee Also
bearing in mind that the leading contributions in the considered
expansion, as $i\lambda \to \infty$, are coming from $Z_{NN}$ and
$\hat Z_{ii}$ for $i \in \{2, \ldots, N \}$ we may write: \be
Z_{NN}(0,-L,\lambda)+ \hat Z_{ii}(0,-L, \lambda) = i \lambda L +
\sum_{n=1}^{\infty} {Z_{NN}^{(n)}(0,-L) +(-1)^n Z_{\bar i \bar
i}^{(n)}(0,-L) \over \lambda^n} \ee Gathering all the information
given above we end up with the following expression \be &&  \ln \
tr \Big \{  T(0,-L ,\lambda)  VT^t(0, -L, -\lambda) V \Big \} =
i\lambda L + \sum_{n=1}^{\infty} {Z^{(n)}_{NN}(0,-L)+(-1)^n
Z_{11}^{(n)}(0-L) \over \lambda^n} \non\\ && +\ln \Big ( 1
+H_{NN}(\lambda) +\bar H_{NN}(\lambda) + \sum_{i=2}^N e^{\hat
Z_{ii}(0,-L,\lambda) -\hat Z_{NN}(0,-L, \lambda)} H_{iN}(\lambda)
\bar H_{Ni}(\lambda) \Big ) \label{fexpb} \ee Finally taking into
account the information provided above
(\ref{nots0})--(\ref{fexpb}) we may express the first non-trivial
integrals of motion as: \be  I_1 &=& Z_{NN}^{(1)}(0,-L)
-Z_{11}^{(1)}(0,-L) \non\\ &=& -i\kappa
\sum_{i=1}^{N-1}\int_{-L}^0 dx\ \psi_{i}(x) \bar \psi_i(x)
-i\kappa \int_{-L}^0 dx\ \psi_{1}(x) \bar \psi_1(x)\label{fnl}\ee
Notice that in general due to the presence of $Z_{11}^{(n)}$ in
(\ref{fexpb}) non-local terms seem to arise in the higher
integrals of motion, which is quite an unusual issue and shall be
addressed elsewhere. Nevertheless, a straightforward computation
of the higher charges, based on the explicit expression
(\ref{fexpb}), may prove the locality or not of the higher
integrals of motions. Moreover, the presence of $Z_{ii}^{(n)}$
alters the structure of the bulk part of the integrals as well.
The latter integral of motion (\ref{fnl}) gives rise to a
`modified' number of particles, ${\cal N}_m =-{I_1 \over
i\kappa}$.

\section{Appendix}

In this appendix we shall evaluate the first integrals of motion
of the quantum discrete $gl_N$ NLS model with SNP boundary
conditions. This model may be also regarded as a higher rank
algebraic extension of the $sl_2$ DST model (see e.g.
\cite{sklyaninblac}), holding a special place between the $gl_N$
quantum spin chains --extensions of the Heisenberg model-- and the
$gl_N$ generalization of the Toda chain. To explicitly specify the
local integrals of motions of the model with open boundary
conditions we shall, as usual, consider the asymptotic expansion
of the generating function ${\cal T}(\lambda)$. We shall focus
here on the simple case where both left and right boundaries are
given by $K^{\pm}(\lambda)= \mbox{antid}(1, \ldots, 1)$, and
effectively we shall expand \be t(\lambda)= tr\ T(\lambda)\ \hat
T(\lambda) ~~~~\mbox{where} ~~~~\hat T(\lambda) = T^t(-\lambda)
\label{ttr} \ee and recall $T(\lambda)$ is given by (\ref{t2}).
Indeed after expanding in powers of $\lambda^{-1}$ we obtain \be
&&T(\lambda) \propto E_{11} + {1 \over \lambda} T^{(1)} + {1 \over
\lambda^2} T^{(2)} + {\cal O}(\lambda^{-3}) \non\\ && \hat
T(\lambda) \propto E_{11} + {1 \over \lambda} \hat T^{(1)} + {1
\over \lambda^2} \hat T^{(2)} + {\cal O}(\lambda^{-3}), \ee where
the quantities $T^{(1,2)},\ \hat T^{(1,2)}$ are defined below \be
T^{(1)} &=& i\kappa \Big (\sum_{n=1}^L {\mathbb N}_n E_{11} +
\sum_{j=2}^{N} \phi_{1}^{(j-1)} E_{1j} +\sum_{j=2}^N
\psi_L^{(j-1)}E_{j1} \Big ) \non\\ \hat T^{(1)} &=&  -i\kappa \Big
(\sum_{n=1}^L \hat {\mathbb N}_n E_{11} + \sum_{j=2}^{N}
\phi_{1}^{(j-1)} E_{j1} +\sum_{j=2}^N \psi_L^{(j-1)}E_{1j}\Big )
\non\\ T^{(2)}&=& -\kappa^2 \Big (\sum_{n>m}{\mathbb N}_{n}
{\mathbb N}_m E_{11}+ \sum_{n=1}^{L-1}\sum_{j=2}^{N}
\psi_n^{(j-1)}\phi_{n+1}^{(j-1)} E_{11} + \sum_{j=2}^{N}
\psi_L^{(j-1)}\phi_{1}^{(j-1)} E_{jj} \non\\ &+&\ \sum_{n=1}^{L-1}
{\mathbb N}_n \sum_{j=2}^N \psi_{L}^{(j-1)}E_{j1} + \sum_{j=2}^N
\phi_1^{(j-1)} \sum_{n=2}^L{\mathbb N}_n E_{1j}
+\sum_{j=2}^{N}\psi_{L-1}^{(j-1)} E_{j1} + \sum_{j=2}^N
\phi_2^{(j-1)} E_{1j}  \Big) \non\\ \hat T^{(2)} &=& -\kappa^2
\Big (\sum_{n<m}\hat {\mathbb N}_{n} \hat {\mathbb N}_m E_{11}+
\sum_{n=1}^{L-1}\sum_{j=2}^{N} \psi_n^{(j-1)}\phi_{n+1}^{(j-1)}
E_{11} + \sum_{j=2}^{N} \psi_L^{(j-1)}\phi_{1}^{(j-1)} E_{jj}
\non\\ &+&\ \sum_{n=1}^{L-1} \hat {\mathbb N}_n \sum_{j=2}^N
\psi_{L}^{(j-1)}E_{1j} + \sum_{j=2}^N \phi_1^{(j-1)} \sum_{n=2}^L
\hat {\mathbb N}_n E_{j1} +\sum_{j=2}^{N}\psi_{L-1}^{(j-1)} E_{1j}
+ \sum_{j=2}^N \phi_2^{(j-1)} E_{j1}  \Big ) \non\\ &&
\mbox{where} ~~~~{\mathbb N}_n = \sum_{j=1}^{N-1} \phi_n^{(j)}
\psi_{n}^{(j)}, ~~~\hat {\mathbb N}_n = {\mathbb N}_n + \rho. \ee
In the expressions above all the lower indices denote the site of
the spin chain, while the upper indices denote the component of
the $(N-1)$ dimensional vector fields.

We may easily obtain first the bulk integrals of motion by
considering the expansion $tr\ T(\lambda) = \sum_{n}{I_n \over
\lambda^n}$. Indeed for the bulk case after simply taking the
trace of $T^{(1, 2)}$ we obtain \be I_1 = i\kappa \sum_{n=1}^L
{\mathbb N}_n , ~~~~I_2 = -\kappa^{2}(\sum_{n<m} {\mathbb N}_n
{\mathbb N}_m + \sum_{n=1}^{L-1}\sum_{j=1}^{N-1} \psi_{n}^{(j)}
\phi_{n+1}^{(j)}+\sum_{j=1}^{N-1} \psi_{L}^{(j)} \phi_{1}^{(j} ).
\label{p1} \ee The quantities identified as the number of
particles and the momentum in the NLS model are given by the
following expressions \be {\cal N}_d = {1 \over i\kappa} I_1,
~~~{\cal P}_d = {1 \over i\kappa} \Big ({1\over 2} I_1^2 - I_2
\Big )\ee and more precisely \be {\cal N}_d= \sum_{n=1}^{L}
{\mathbb N}_n, ~~~~~{\cal P}_d = -{ i \kappa \over 2} \Big
(\sum_{n=1}^{L} {\mathbb N}_n^2
-2\sum_{n=1}^{L-1}\sum_{j=1}^{N-1}\psi_{n}^{(j)} \phi_{n+1}^{(j)}
- 2\sum_{j=1}^{N-1} \phi_1^{(j)} \psi_{L}^{(j)} \Big ).
\label{bulkd} \ee We come now to the open NLS model, and we
consider the expansion of (\ref{ttr}). The first charge of the
open model is zero, that is the number of particles is not a
conserved quantity anymore. The second charge is given by \be I_2=
\kappa^2 \Big (\sum_{n=1}^L {\mathbb N}_n^{2} - 2
\sum_{n=1}^{L-1}\sum_{j=1}^{N-1} \psi_n^{(j)} \phi_{n+1}^{(j)} +
\sum_{j=1}^{N-1}
(\psi_L^{(j)}\psi_L^{(j)}+\phi_1^{(j)}\phi_1^{(j)})  \Big )
\label{p2} \ee and corresponds to the momentum ${\cal P}_d = {I_2
\over 2 i\kappa}$, which is obviously modified due to the presence
of the open boundaries. The third charge again is trivial,
involving only boundary terms.  We do not compute any higher
conserved charges, but we may rather safely conjecture that the
only non-trivial conserved charges are the even ones.

\end{document}